\documentclass[12pt]{article}

\begin{document}

\begin{center}
{\bf Generalized Maxwell Equations and Their Solutions } \\
\vspace{5mm} S.I.Kruglov \footnote{E-mail:
skruglov23@hotmail.com}\\
 \vspace{3mm} \textit{International
Educational Centre, 2727 Steeles Ave.West, Suite 202,
\\Toronto, Ontario, Canada M3J 3G9} \\

\vspace{5mm}
\end{center}

\begin{abstract}
    The generalized Maxwell equations are considered which include
an additional gradient term. Such equations describe massless particles
possessing spins one and zero. We find and investigate the matrix formulation
of the first order of equations under consideration. All the linearly
independent solutions of the equations for a free particle are obtained
in terms of the projection matrix-dyads (density matrices).
\end{abstract}

\section{Introduction}
Now the Dirac-K\"ahler field in the framework of differential
forms is of interest [1]. This is due to the possibility of using
the Dirac-K\"ahler equation for describing fermions with spin
$1/2$ on the lattice [2]. K\"ahler [1] investigated an equation
for inhomogeneous differential forms
\begin{equation} \left(
d-\delta +m\right) \Phi =0,  \label{1}
\end{equation}
where $m$ is the mass, $d$ being the exterior derivative, $\delta
=-\star^{-1} d\star $ turns $n- $forms into $(n-1)-$form; the
$\star $ is the operator connecting a $n-$form with a
$(4-n)-$form; $\star ^2=1$, $d^2=\delta ^2=0 $; $\left( d-\delta
\right) ^2=-\left( d\delta +\delta d\right) =\partial _\mu^2$,
$\partial _\mu=\partial/\partial x_\mu=(\partial/\partial
x_m,\partial/i\partial t$, $t$ is the time. The inhomogeneous
differential form $\Phi$ is given by
\[
\Phi =\varphi (x)+\varphi _\mu (x)dx^\mu +\frac 1{2!}\varphi _{\mu
\nu }(x)dx^\mu \wedge dx^\nu +
\]
\vspace{-8mm}
\begin{equation}
\label{2}
\end{equation}
\vspace{-8mm}
\[
+\frac 1{3!}\varphi _{\mu \nu \rho }(x)dx^\mu \wedge dx^\nu \wedge
dx^\rho +\frac 1{4!}\varphi _{\mu \nu \rho \sigma }(x)dx^\mu
\wedge dx^\nu \wedge dx^\rho \wedge dx^\sigma
\]
 what is equivalent to introducing a scalar field $\varphi (x)$,
 vector field $\varphi _\mu (x)$, and antisymmetric tensor fields:
$\varphi _{\mu \nu }(x)$, $\varphi _{\mu \nu \rho }(x)$, $\varphi
_{\mu \nu \rho \sigma }(x)$. The antisymmetric tensors $ \varphi
_{\mu \nu \rho }(x)$, $\varphi _{\mu \nu \rho \sigma }(x)$ define
a pseudovector and pseudoscalar fields, respectively:
\begin{equation}
\widetilde{\varphi }_\mu (x)=\frac 1{3!}\varepsilon _{\mu \nu \rho
\sigma }\varphi _{\nu \rho \sigma
}(x),\hspace{0.5in}\widetilde{\varphi }(x)=\frac 1{4!}\varepsilon
_{\mu \nu \rho \sigma }\varphi _{\mu \nu \rho \sigma }(x),
\label{3}
\end{equation}
where $\varepsilon _{\mu \nu \alpha \beta }$ is an antisymmetric
tensor Levy-Civita; $\varepsilon _{1234}=-i$. So, Eq.(1) includes
two scalar and two vector fields. This means the consideration of
fields with spins zero and one with the same mass $m$. Eq.(1) with
the definitions (2), (3) can be represented as the following
system of tensor fields [3]:
\[
\partial _\nu \psi _{[\mu \nu] }(x)-\partial _\mu \psi (x)+m^2
\psi_\mu (x)=0, \hspace{0.5in}\partial _\nu \widetilde{\psi
}_{[\mu \nu] }(x)-\partial _\mu \widetilde{\psi
}(x)+m^2\widetilde{\psi}_\mu (x)=0,
\]
\begin{equation}
\partial _\mu \psi_\mu (x)=\psi (x),\hspace{0.3in}\partial _\mu
\widetilde{\psi}_\mu (x)=\widetilde{\psi }(x),  \label{4}
\end{equation}
\[
\psi _{[\mu \nu ]}(x)=\partial _\mu \psi_\nu (x)-\partial _\nu
\psi_\mu (x)-\varepsilon _{\mu \nu \alpha \beta }\partial _\alpha
\widetilde{\psi}_\beta (x),
\]

where $\widetilde{\psi }_{[\mu \nu] }=(1/2)\varepsilon _{\mu \nu
\alpha \beta }\psi _{\alpha \beta }$ is the dual tensor. There is
the doubling of the spin states of fields described because
Eqs.(4) contain two four-vectors $\psi_\mu(x)$, $\widetilde{\psi
}_\mu(x)$ and two scalars $\psi(x)$, $\widetilde{\psi}(x)$).
Equations (4) can be represented as the $16$ -dimensional first
order Dirac equation [3]. That is why there is a connection
between description of fermions with spin $1/2$ and bosonic fields
$\psi(x) $, $\psi_\mu(x)$, $\psi _{\mu \nu }(x)$, $\widetilde{\psi
}(x)$, $\widetilde{\psi}_\mu(x)$. At the restrictions
$\widetilde{\psi}_\mu=0$, $\widetilde{\psi}=0$ we arrive at the
Proca equations [4]. Stueckelberg's equation [5], describing
fields with spin one and zero, corresponds to the case
$\widetilde{\psi}_\mu=0$ in (4).

From Eqs.(4) at $m=0$ we arrive at the two-potential formulation
of massless fields with two gradient terms
\[
\partial _\nu \psi _{[\mu \nu] }(x)-\partial _\mu \psi (x)
=0, \hspace{0.5in}\partial _\nu \widetilde{\psi }_{[\mu \nu]
}(x)-\partial _\mu \widetilde{\psi }(x)=0,
\]
\begin{equation}
\partial _\mu \psi_\mu (x)=\psi (x),\hspace{0.3in}\partial _\mu
\widetilde{\psi}_\mu (x)=\widetilde{\psi }(x),  \label{5}
\end{equation}
\[
\psi _{[\mu \nu ]}(x)=\partial _\mu \psi_\nu (x)-\partial _\nu
\psi_\mu (x)-\varepsilon _{\mu \nu \alpha \beta }\partial _\alpha
\widetilde{\psi}_\beta (x).
\]
Eqs.5 represent the generalized Maxwell equations which were
studied in [6-11]. In [3] we found and investigated the matrix
formulation of the first order of equations (5) and solutions for
a free particle in the form of the projection matrix-dyads. The
matrices of an equation obey the Dirac algebra. In this work we
study ``minimal" generalization of Maxwell's equations by setting
$\widetilde{\psi}_\beta (x)=\widetilde{\psi}(x)=0$ in (5). In this
case there is no doubling of spin states of fields: there is one
state with spin zero and two states with helicity $\pm 1$.

\section{Matrix form of massless bosonic fields}

Let us consider the following generalized Maxwell equations (see
also [12-14])
\[
\partial _\nu \psi _{[\mu \nu ]}+\partial _\mu \psi _0=0,
\]
\begin{equation}
\partial _\nu \psi _\mu -\partial _\mu \psi _\nu +\kappa\psi _{[\mu \nu ]}=0,
\end{equation}
\[
\partial _\mu \psi _\mu +\kappa\psi _0=0.  \label{6}
\]

Eqs.(6) follow from (5) at the replacement $\widetilde{\psi}_\beta
(x)=\widetilde{\psi}(x)=0$, $\psi_{\mu\nu}\rightarrow\kappa
\psi_{\mu\nu}$, $\psi (x)\rightarrow -\kappa \psi_0 (x)$. Fields
$\psi _\mu $, $\psi _0$ are massless vector and scalar fields,
respectively, and $\kappa$ is a parameter which we introduced for
convenience. So, equations (6) describe massless particles
possessing spins one and zero without doubling of spin states. The
classical Maxwell equations are obtained by setting $\psi_0=0$.

It is easy to get the massive fields by adding the term
$m\psi_\mu$ in the first equation (6) (at $\kappa=m$). In this
case we arrive at the massive (with mass $m$) Stueckelberg fields
[5,15]. In [15] the matrix form of equations for massive fields
and solutions in the form of the projection matrix-dyads were
found.

Now we consider the matrix formulation of the first order of the
field equations (6) for massless fields which is convenient for
constructing the density matrix and for some electrodynamics
calculations. Let us introduce the matrix $\varepsilon ^{A,B}$
[16] with dimension $ n\times n$; its elements consist of zeroes
and only one element is unity where row $A$ and column $B$ cross.
So the matrix elements and multiplication of these matrices are
\begin{equation}
\left( \varepsilon ^{A,B}\right)_{CD}=\delta_{AC}\delta_{BD},
\hspace{0.5in}\varepsilon ^{A,B}\varepsilon ^{C,D}=\delta
_{BC}\varepsilon ^{A,D},  \label{7}
\end{equation}
where indexes $A,B,C,D=1,2,...n$.
After introducing the 11-dimensional function
\begin{equation}
\Psi (x)=\left\{ \psi _A(x)\right\} =\left(
\begin{array}{c}
\psi _0 \\
\psi _\mu \\
\psi _{[\mu \nu ]}
\end{array}
\right) \hspace{0.5in}(A=0,\mu,[\mu \nu ]),  \label{8}
\end{equation}
where $\mu ,$ $\nu =1,$ $2,$ $3,$ $4$, and using the elements of
the entire algebra (7), Eq.(6) can be written in the form of one
equation
\[
\partial _\nu \left( \varepsilon ^{\mu ,[\mu \nu ]}+\varepsilon ^{[\mu \nu
],\mu }+\varepsilon ^{\nu ,0}+\varepsilon ^{0,\nu }\right) _{AB}\psi _B(x)+
\]
\vspace{-8mm}
\begin{equation}
\label{9}
\end{equation}
\vspace{-8mm}
\[
+\kappa\left( \varepsilon ^{0,0}+\frac 12\varepsilon ^{[\mu \nu
],[\mu \nu ]}\right)_{AB}\psi _B(x)=0.
\]

Introducing 11-dimensional matrices
\[
\alpha _\nu =\varepsilon ^{\mu ,[\mu \nu ]}+\varepsilon ^{[\mu \nu ],\mu
}+\varepsilon ^{\nu ,0}+\varepsilon ^{0,\nu },
\]
\vspace{-8mm}
\begin{equation}
\label{10}
\end{equation}
\vspace{-8mm}
\[
\overline{P}=\varepsilon ^{\mu ,\mu },\hspace{0.5in}P=\varepsilon
^{0,0}+\frac 12\varepsilon ^{[\mu \nu ],[\mu \nu ]},
\]

Eq.(9) takes the form of the relativistic wave equation of the
first order:
\begin{equation}
\left( \alpha _\mu \partial _\mu +\kappa P\right) \Psi (x)=0.
\label{11}
\end{equation}

So, matrix equation (11) gives a unified description of a scalar
and vector massless fields.

Matrices $\overline{P}$, $P$ are the projective matrices (see
[16,17]) which obey the relations:
\[
P^2=P,\hspace{0.5in}\overline{P}^2=\overline{P},\hspace{0.5in}P+\overline{P}
=I_{11},
\]
\vspace{-8mm}
\begin{equation}
\label{12}
\end{equation}
\vspace{-8mm}
\[
\alpha _\mu \overline{P}+\overline{P}\alpha _\mu =\alpha _\mu
,\hspace{0.5in} \alpha _\mu P+P\alpha _\mu =\alpha _\mu ,
\]
where $I_{11}$ is the unit matrix in $11-$dimensional space. The
Stueckelberg equation for massive fields in the matrix form is
given by [15].

\begin{equation}
\left( \alpha _\mu \partial _\mu +m\right) \Psi (x)=0.  \label{13}
\end{equation}

It should be noted that the matrices $\alpha _\mu $ can be represented as
\[
\alpha _\mu =\beta _\mu ^{(1)}+\beta _\mu ^{(0)},
\]
\begin{equation}
\beta _\nu ^{(1)}=\varepsilon ^{\mu ,[\mu \nu ]}+\varepsilon ^{[\mu \nu
],\mu },
\end{equation}\label{14}
\[
\beta _\nu ^{(0)}=\varepsilon ^{\nu ,0}+\varepsilon ^{0,\nu },
\]

where the $10-$dimensional $\beta _\mu ^{(1)}$ and $5-$dimensional
$\beta _\mu ^{(0)}$ matrices obey the Petiau-Duffin-Kemmer [18-20]
algebra:
\begin{equation}
\beta _\mu \beta _\nu \beta _\alpha +\beta _\alpha \beta _\nu
\beta _\mu =\delta _{\mu \nu }\beta _\alpha +\delta _{\alpha \nu
}\beta _\mu , \label{15}
\end{equation}

so that the equations for massive spin-$1$ and spin-$0$ particles
are (see [16])
\begin{equation}
\left( \beta _\mu ^{(1)}\partial _\mu +m\right) \Psi ^{(1)}(x)=0,
\hspace{0.5in}\Psi ^{(1)}(x)=\left(
\begin{array}{c}
\psi _\mu \\
\psi _{[\mu \nu ]}
\end{array}
\right),  \label{16}
\end{equation}
\begin{equation}
\left( \beta _\mu ^{(0)}\partial _\mu +m\right) \Psi ^{(0)}(x)=0,
\hspace{0.5in}\Psi ^{(0)}(x)=\left(
\begin{array}{c}
\psi _0 \\
\psi _\mu
\end{array}
\right).  \label{17}
\end{equation}

The $10-$dimensional Petiau-Duffin-Kemmer equation (16) is
equivalent to the Proca equations [4] for spin-$1$ particles and
the $5-$dimensional Eq. (17) is equivalent to the
Klein-Gordon-Fock equation for scalar particles. The
$11-$dimensional Eq.(11) describes massless fields with two spins
$0,$ $1$ (multi-spin 0,1). It is not difficult to verify (using
Eqs.(7)) that the $11-$dimensional matrices $\alpha_\mu$ (10)
satisfy the algebra (see also [21]):
\[
\alpha _\mu \alpha _\nu \alpha _\alpha +\alpha _\alpha \alpha _\nu \alpha
_\mu +\alpha _\mu \alpha _\alpha \alpha _\nu +\alpha _\nu \alpha _\alpha
\alpha _\mu +\alpha _\nu \alpha _\mu \alpha _\alpha +\alpha _\alpha \alpha
_\mu \alpha _\nu =
\]
\vspace{-8mm}
\begin{equation}
\label{18}
\end{equation}
\vspace{-8mm}
\[
 =2\left( \delta _{\mu \nu }\alpha _\alpha +\delta
_{\alpha \nu }\alpha _\mu +\delta _{\mu \alpha }\alpha _\nu
\right).
\]

This algebra is more complicated than the Petiau-Duffin-Kemmer
algebra (15). Different representations of the
Petiau-Duffin-Kemmer algebra (15) were considered in [22-27].

\section{Solutions of generalized Maxwell's equations}

Let us now consider the solutions of the matrix equation (11) for
massless fields. In the momentum space, Eq.(11) is given by
\begin{equation}
D\Psi _k=0,\hspace{0.5in}D=i\widehat{k}+\kappa P,  \label{19}
\end{equation}

where $\widehat{k}=\alpha _\mu k_\mu $, $k_\mu ^2={\bf k}^2-k_0^2=0$ and
the matrix $D$ obeys the minimal equation
\begin{equation}
D\left( D-\kappa\right) ^2=0.  \label{20}
\end{equation}

It should be noted that this matrix equation of generalized
Maxwell's equations with multi-spin $0,1$ is simpler than the
minimal equation for Maxwell's equations with pure spin $1$
[16,28]. Using the general scheme [17] we find that the projection
operator corresponding to eigenvalue $0$ of the operator $D$ is
\begin{equation}
\gamma =\left( \frac{D-\kappa}{\kappa}\right) ^2,  \label{21}
\end{equation}

so that $\gamma ^2=\gamma $.

Every column of the matrix $\gamma $ can be considered as an
eigenvector $\Psi_k$ of equation (19) with eigenvalue $ 0$.
Eq.(19) for projection operators tells that matrix $ \gamma $ can
be transformed into diagonal form, with the diagonal containing
only ones and zeroes. So the $\gamma $ acting on any function
$\Psi $ will retain components which are solutions of Eq.(19).

The generators of the Lorentz group in the $11-$dimensional
space being considered are given by
\begin{equation}
J_{\mu \nu }=\beta _\mu ^{(1)}\beta _\nu ^{(1)}-\beta _\nu
^{(1)}\beta _\mu ^{(1)}.  \label{22}
\end{equation}

It should be noted that matrices (22) act in the $10-$dimensional
subspace $\left( \psi _\mu ,\psi _{[\mu \nu ]}\right) $ because
the scalar $ \psi _0$ is an invariant of the Lorentz
transformations. So matrices (22) are also generators of the
Lorentz group for the Petiau-Duffin-Kemmer fields of Eq.(16).
Using properties (7), we get the commutation relations
\begin{equation}
\left[ J_{\rho \sigma },J_{\mu \nu }\right] =\delta _{\sigma \mu
}J_{\rho \nu }+\delta _{\rho \nu }J_{\sigma \mu }-\delta _{\rho
\mu }J_{\sigma \nu }-\delta _{\sigma \nu }J_{\rho \mu },
\label{23}
\end{equation}
\begin{equation}
\left[ \alpha _\lambda ,J_{\mu \nu }\right] =\delta _{\lambda \mu
}\alpha _\nu -\delta _{\lambda \nu }\alpha _\mu .  \label{24}
\end{equation}

Relationship (23) is a well known commutation relation for
generators of the Lorentz group $SO(3,1)$. Equation (11) is
form-invariant under the Lorentz transformations since relation
(24) is valid. To guarantee the existence of a relativistically
invariant bilinear form
\begin{equation}
\overline{\Psi }\Psi =\Psi ^{+}\eta \Psi,  \label{25}
\end{equation}

where $\Psi ^{+}$ is the Hermitian-conjugate wave function, we
should construct a Hermitianizing matrix $\eta $ with the
properties [16,17,24]:
\begin{equation}
\eta \alpha _i=-\alpha _i\eta ,\hspace{0.5in}\eta \alpha _4=\alpha
_4\eta \hspace{0.5in}(i=1,2,3).  \label{26}
\end{equation}

Such a matrix exists and is given by
\[
\eta =-\varepsilon ^{0,0}+2\beta _4^{(1)2}-I_{10},
\]
\vspace{-8mm}
\begin{equation}
\label{27}
\end{equation}
\vspace{-8mm}
\[
 I_{10}=\varepsilon ^{\mu ,\mu }+\frac 12\varepsilon
^{[\mu \nu ],[\mu \nu ]},
\]

where the matrix $\eta ^{(1)}=2\beta _4^{(1)2}-I_{10}$ plays the
role of a Hermitianizing matrix for the Petiau-Duffin-Kemmer
equation (16) [16]. The operator of the squared spin (squared
Pauli-Lubanski vector) is given by
\begin{equation}
\sigma ^2=\left( \frac 1{2k_0}\varepsilon _{\mu \nu \alpha \beta
}k_\nu J_{\alpha \beta }\right) ^2=\frac 1{k_0^2} J_{\mu \sigma
}J_{\sigma \nu}k_\mu k_\nu.  \label{28}
\end{equation}

It may be verified that this operator obeys the minimal equation
\begin{equation}
\sigma ^2\left( \sigma ^2-2\right) =0,  \label{29}
\end{equation}

so that eigenvalues of the squared spin operator $\sigma ^2$ are $s(s+1)=0$
and $s(s+1)=2$. This confirms that the considered fields describe the
superposition of two spins $s=0$ and $s=1$. To separate these states we use
the projection operators
\begin{equation}
S_{(0)}^2=1-\frac{\sigma ^2}2,\hspace{0.5in}S_{(1)}^2=\frac{\sigma
^2}2 \label{30}
\end{equation}

having the properties $S_{(0)}^2S_{(1)}^2=0$, $\left( S_{(0)}^2\right)
^2=S_{(0)}^2$, $\left( S_{(1)}^2\right) ^2=S_{(1)}^2$, $
S_{(0)}^2+S_{(1)}^2=1 $, where $1\equiv I_{11}$ is the unit matrix in $11-$
dimensional space. In accordance with the general properties of the
projection operators, the matrices $S_{(0)}^2$, $S_{(1)}^2$ acting on the
wave function extract pure states with spin $0$ and $1$, respectively. Now
we introduce the operator of the spin projection on the direction of the
momentum ${\bf k}$ (helicity) :
\begin{equation}
\sigma _k=-\frac i{2k_0}\epsilon _{abc}k_aJ_{bc}=-\frac
i{k_0}\epsilon _{abc}k_a\beta _b^{(1)}\beta _c^{(1)}.  \label{31}
\end{equation}
The minimal matrix equation for the spin projection operator is
\begin{equation}
\sigma_k\left( \sigma_k-1\right) \left( \sigma_k+1\right) =0
\label{32}
\end{equation}

and the corresponding projection operators are given by
\begin{equation}
\widehat{S}_{(\pm 1)}=\frac 12\sigma_k\left( \sigma_k\pm 1\right),
\hspace{0.5in}\widehat{S}_{(0)}=1-\sigma_k^2.  \label{33}
\end{equation}

Operators $\widehat{S}_{(\pm 1)}$ correspond to the spin
projections $s_k=\pm 1$. It is easy to verify that the required
commutation relations hold:
\[
\left[ S_{(0)}^2,\widehat{k}\right] =\left[ S_{(1)}^2,\widehat{k}\right]
=\left[ \widehat{S}_{(\pm 1)},\widehat{k}\right] =\left[ \widehat{S}_{(0)},
\widehat{k}\right] =0,
\]
\vspace{-8mm}
\begin{equation}
\label{34}
\end{equation}
\vspace{-8mm}
\[
\left[ S_{(0)}^2,\widehat{S}_{(\pm 1)}\right]
=\left[ S_{(1)}^2,\widehat{S} _{(\pm 1)}\right] =\left[
S_{(0)}^2,\widehat{S}_{(0)}\right] =0.
\]

Thus the projection matrices extracting pure states with definite spin (0 and 1),
and spin projections (helicity $\pm 1$) take the form
\[
\Pi _{(0)}=\left( 1-\frac{\sigma ^2}2\right) \left( \frac{D-\kappa}{\kappa}\right)
^2,
\]
\vspace{-8mm}
\begin{equation}
\label{35}
\end{equation}
\vspace{-8mm}
\[
 \Pi _{(\pm 1)}=\frac 12\sigma _k\left( \sigma _k\pm
1\right) \left( \frac{ D-\kappa}{\kappa}\right) ^2,
\]
where we took into account that $\left( \sigma ^2/2\right) \sigma
_k=\sigma _k$. Projection operators $\Pi _{(0)}$, $\Pi _{(\pm 1)}$
extract states with spin $0$ and $1$, respectively. The $\Pi
_{(0)}$, $\Pi _{(\pm 1)}$  are the density matrices for pure spin
spates. It is easy to consider impure states by summation of
Eqs.(35) over spin projections and spins. Projection operators for
pure states can be represented as matrices-dyads [17]:
\begin{equation}
\Pi_{(0)}=\Psi _{(0)}\cdot \overline{\Psi }_{(0)},\hspace{0.5in}
\Pi_{(\pm 1)}=\Psi_{(\pm )}\cdot \overline{\Psi }_{(\pm )}, \label{31}
\end{equation}

where the wave functions $\Psi_{(0)}$, $\Psi_{(\pm )}$ correspond
to spin $0$ and $1$, respectively. Solutions of Eq.(13) for
massive particles with spins 0 and 1 in the form of matrix-dyads
were found in [15].

Expressions (35), (36) are convenient for calculating different
electrodynamics processes involving polarized massless particles.
It is possible to make evaluations of different physical
quantities in a covariant manner without using the matrices of
first-order equations in a definite representation.

\section{Conclusion}

Compared to the Maxwell equations which describe left and right
polarized waves (helicity $\pm1$), Eqs.(6) admit also an
additional longitudinal state corresponding to spin-zero of the
field. This state gives the negative contribution to the
Hamiltonian of fields under consideration and it is necessary to
introduce an indefinite metric to quantize such a field (see
[15]). To eliminate the additional state with spin-zero one may
impose the constraint $\psi_0 (x)=0$ in equations (6), and we
arrive at the classical Maxwell equations, where
$\psi_{\mu\nu}(x)$ is the strength tensor; $E_m=i\psi_{m4}$,
$H_m=(1/2)\varepsilon _{m n k}\psi_{n k}$ are electric and
magnetic fields, respectively. It is possible also to treat the
scalar field $\psi_0 (x)$ as non-physical one in the general gauge
$\psi_0 (x)\neq 0$ (an orthodox point of view). In this way after
some calculations one should eliminate the contribution of this
non-physical scalar field in this general gauge. In extraordinary
point of view, vector and scalar states of the system (6) can be
treated on the same footing with introducing indefinite metric.
This, however, requires the further development and physical
interpretation of quantum field theory with indefinite metric.

\end{document}